\documentclass[preprint,authoryear,12pt]{elsarticle}
\pdfoutput=1 

\usepackage{graphicx}

\usepackage{amssymb}

\journal{Icarus}

\begin{document}

\begin{frontmatter}


\ead{nicole.meyer@obspm.fr}
\ead[url]{http://www.lesia.obspm.fr/perso/nicole-meyer/}

\title{On the charge of nanograins in cold environments\\  and Enceladus dust}


\author{N. Meyer-Vernet}

\address{LESIA, Observatoire de Paris - CNRS - Universit\'{e} Pierre et Marie Curie -
Universit\'{e} Denis Diderot, Meudon, France.}

\begin{abstract}
In very-low energy   plasmas, the size of nanograins is comparable to  the distance (the so-called Landau length) at which the interaction energy of two electrons equals  their thermal energy. In that case, the grain's polarization induced by  approaching charged particles  increases their fluxes and reduces the charging time scales. Furthermore,  for grains of radius smaller than the Landau length,  the  electric charge  no longer decreases linearly with size, but has a most probable equilibrium value  close to one electron charge.  We give analytical results that can be used for  nanograins in cold dense planetary environments of the outer solar system. Application to the nanodust observed in the plume of Saturn's moon Enceladus shows that most grains of radius about 1 nm should carry one electron, whereas an appreciable fraction of them are positively charged by ion impacts. The corresponding electrostatic stresses should destroy smaller grains, which anyway may not exist as crystals since their number of molecules is close to the minimum required for crystallization.

\end{abstract}


\end{frontmatter}


\section{Introduction}

Dust particles of nanometric size have been detected in situ in various parts of the solar system, e.g. near comets  \citep{utt90}, in the Earth low ionosphere (e.g. \cite{fri09} and references therein), in the solar wind near 1 AU  \citep{mey09}, in streams ejected by Jupiter \citep{zoo96} and Saturn \citep{kem05}, in the atmosphere of Saturn's moon Titan \citep{coa07}, and in the plume ejected by the icy moon  Enceladus  \citep{jon09}. Nanograins, which make the transition between molecules and bulk materials, can be produced by condensation of gases and aggregation of molecules \citep{kim12},   and/or by fragmentation of larger dust (e.g. \citep{man12}). The  large surface-to-volume ratio of nanoparticles makes the proportion of surface atoms significant, so that their characteristic properties often differ from those of bulk materials and they are major agents for interactions with particles and fields.

Nanograins play an important role in magnetized environments because their interaction with electromagnetic fields varies in proportion of their electric charge, which varies more slowly with size than do the friction forces (proportional to surface) and the gravitational forces (proportional to volume). Hence nanograins are generally driven by electromagnetic forces as are plasma particles, so that their electric charge governs their dynamics (e.g. \cite{bur01,hor96,man12,man13}). The electric charge can also determine the grain's minimum size via the electrostatic stresses producing fracture, and it also affects the grains' growth and coalescence. At larger scales, it determines  the Larmor frequency and thus the time scale of grains' pick-up.

At nanometric sizes, several effects make the charging processes different from the classical  charging of larger objects. First, it is well known that  the particle sticking coefficients and photoelectric and secondary emission yields can change  \citep{wat72,cho93,wei01,abb10}, essentially because the electron free path in matter is of the order of (or larger than) 1 nm below $\sim 10$ eV \citep{fit01}.

Two further effects appear when the grain radius becomes comparable to
\begin{equation}
r_L = e^2/(4 \pi \epsilon _0 k_B T)
\end{equation}
in a plasma of temperature $T$.  Since $r_{L{\rm (nm)}} \simeq 1.44/T_{{\rm eV}}$, this concerns   nanograins in plasmas of temperature $\simeq 1$ eV.  This scale, often called  the Landau radius, is the distance below which the mutual electrostatic energy of two approaching charged particles exceeds the kinetic energy of their relative motion, so that they significantly perturb each other's trajectories. This fundamental scale, which determines the plasma particle cross-sections for Coulomb collisions producing large perturbations, is also of major importance for dust grains. Indeed,  the  particles approaching a grain of radius $\sim r_L$ or smaller induce polarization charges  whose Coulomb attraction increases the  collected fluxes, thereby decreasing the charging time scales. Furthermore,  since at this scale the charging becomes discretized, the equilibrium charge on a grain  no longer varies in proportion of its size, but  becomes comparable to one electron charge  in a wide range of  sizes, because  the  probability that an uncharged grain collects an electron exceeds  the probability that a neutral or negatively charged grain collects an ion.

The latter phenomena have been studied in the contexts of the Earth's ionosphere (e.g.    \cite{jen91,rap01} and references therein) and of the interstellar medium (e.g. \cite{dra87,wei01}). In this paper, we consider these effects for cold dense planetary environments in the outer solar system, which are subjected to different constraints. We derive analytical results that can be used in these contexts, and apply them to the nanograins detected in  Enceladus plume  \citep{jon09, hil12}, where the electrons are cold  enough  \citep{sha11} to put the Landau radius in the nano range, the plasma is dense enough \citep{mor11} for the photoelectron emission to be negligible, and the (larger) dust concentration is high enough to deplete the electrons by a large amount.

These calculations will enable us  to estimate the grains' size limit set by electrostatic disruption and to compare it with other physical processes.

Units are SI, unless otherwise indicated explicitly.

\section{Basic impact charging \label{chargingimpacts}}

Before considering nanograins, let us briefly summarize the classical electric charging by collection and emission of particles for a  dust grain of radius $a\gg r_L$ in a  plasma whose electron and ion densities may be different because of the possible presence of dust.

\subsection{Impacts of charged particles}

The charging of a grain changes its  electric potential, which changes the particle fluxes until an equilibrium is reached when the different charge fluxes balance each other. 
The electron flux tends to exceed that of ions because of the faster electron speeds (except in the case of strong electron depletion discussed in Sect.~\ref{dusty}); hence, when the charging is mainly due to electron and ion impacts, the body charges negatively  until  it repels sufficiently the electrons for their flux to balance that of positive ions. For this to be so, the electron potential energy at the body's surface $e \Phi$ must exceed sufficiently (but not too much) the particle thermal energy $\sim k_BT$. Thus the equilibrium grains' potential with respect to the ambient plasma is $\Phi \simeq - \eta k_B T/e$, with  $\eta$ of order of magnitude unity. For a sphere of radius $a $ much smaller than both the Debye length $L_D$ and  the grains' separation, the  electric charge is $Q = 4 \pi \epsilon _0 a \Phi  $. Substituting the above value of $\Phi $ yields the number of charge units $Q/e$ at equilibrium
\begin{equation}
Z = - \eta a/r_L \label{Z1}
\end{equation}
with  $\eta\sim 1$ in  order of magnitude. The mean number of a grain's charges thus exceeds unity when  $a/r_L \gg 1$.

The parameter $\eta$ is  easily calculated  since in that case  the particles are subjected to the Coulomb potential of the grain without intervening barriers of potential (the so-called orbit-limited condition \citep{laf73,whi81}). When the plasma particles are singly charged and have  isotropic Maxwellian velocity distributions, the classical fluxes of each particle species can then be expressed straightforwardly as
\begin{eqnarray}
N& = & N_0 e^{- |\eta |} = N_0 e^{- |Z| r_L/a}  \;\;\;\;\;\;\;\;\;\;\;\;\;\;\;\;\;\; {\rm repelled \;\; particles} \label{Nrep} \\ 
N& = & N_0 (1+  |\eta |) = N_0 ( 1+ |Z| r_L/a)   \;\;\;\; {\rm attracted \;\; particles}   \label{Natt} 
\end{eqnarray} 
per unit grain's surface, where $r_L$ is the Landau radius corresponding to the temperature of the species considered and $N_0$ is the flux of that species on an uncharged grain
\begin{equation}
 N_0 = s\;n \langle v\rangle /4 = s\;n\;(k_BT/2\pi m)^{1/2} \label{N0}
\end{equation}
Here $s$, $n$, $T$, $m$, and $ \langle v\rangle$ are the sticking probability, number density, temperature, mass, and mean speed of the species concerned   in the unperturbed plasma.   For ions of mass $m_i$ and same  temperature $T$ as electrons of mass $m_e$, we have
\begin{equation}
 N_{0e}/N_{0i}= \mu (n_e/n_i) \;\;\;\;\; {\rm with \;\;} \mu =(s_e/s_i)(m_i/m_e)^{1/2}   \label{mu}
\end{equation}
At equilibrium, the electron  and ion fluxes  balance, and $\eta$ is the solution of the equation
\begin{equation}
\eta =   \ln [\mu(n_e/n_i)/(1+\eta)]   \label{Zeq}
\end{equation}

This confirms that $\eta$ is of order of magnitude unity, except if $\mu(n_e/n_i) \simeq 1$ - a case that we will discuss later. 
Note that  we have not assumed $n_e = n_i$, in order for the results to be applicable in dusty environments. Therefore, although $\mu \gg 1$ because of the large ion-to-electron mass ratio, we have not necessarily $\mu(n_e/n_i \gg 1$, but only the weaker inequality $\mu(n_e/n_i) > 1$ (as will be shown in Sect.~\ref{dusty}). 

For nanograins in a low-energy plasma, the sticking probability of ions  $s_i\simeq$ 1, but  the sticking probability of electrons may be smaller because the free path of electrons in solids (which decreases as energy decreases at energies exceeding a few 100 eV) reaches a minimum generally smaller than 1 nm in the vicinity of tens eV, and increases as energy decreases again to values comparable to or greater than 1 nm around 1 eV, taking into account elastic and inelastic scattering  \citep{fit01}. A conservative assumption is $s_e \simeq  0.3-1 $  \citep{jur95,vos06,meg09} for nanograins, keeping in mind that $s_e$ may be much smaller as the number of atoms decreases  \citep{mic87}, essentially because the limited number of degrees of freedom precludes the conservation of energy and momentum in the collision.

For  water-group incident ions $\mu \simeq $  $180\times s_e$, so that  Eq.(\ref{Zeq}) yields $\eta \simeq $ 0.31, 1.85, or 3.65 for respectively $s_e n_e/n_i =0.01$, 0.1,  or 1.  

\subsection{Other charging processes}

The above estimates assume that photoemission (including photodetachment) and secondary emission are negligible. The photoelectron emission on uncharged grains at heliospheric distance $r_{{\rm AU}}$ (in astronomical units) can be approximated by \citep{gra73}
\begin{equation}
N_{ph0}  \simeq  0.5 \times 10^{14} \; \chi /r_{\rm{AU}}^2 \;\;\;\;\;\;\;\;\;\chi  \sim  
0.1-1\label{Nph0}
\end{equation}
per unit of the total grain's surface area $4 \pi a^2$ - to facilitate comparison with  other fluxes (we have taken into account that the projected sunlit area is one-quarter of the grain's surface area). The smaller value of $\chi $ corresponds for example to materials such as graphite or ice, the larger to silicates.

For nanograins $\chi$ may be different for two main reasons which act in opposite senses. First, since the photon attenuation length generally exceeds the photoelectron escape length by a large amount, a small grain size limits the distance from the excitation region to the surface, which tends to increase the yield compared to that of bulk materials  \citep{wat72,wei01}. Second, the photon absorption cross-section (normalized to the cross-sectional area) at the relevant wave lengths $\lambda \sim 0.1 \;\mu$m  varies roughly as  $2 \pi a /\lambda \simeq 6 \times 10^{-2}\; a_{\rm nm}$ when this size parameter is much smaller than unity; this is expected to decrease $\chi $ significantly. Because of the large uncertainties in these properties, we will use the conservative assumption $\chi  \leq 0.1$ for silicate and water ice nanograins. From (\ref{Nph0}), we deduce the ratio between photoelectron emission and ion (of mass $A m_p$) collection for uncharged grains at 10 AU heliocentric distance ($\simeq $ Saturn's orbit), 
\begin{equation}
N_{ph0}/N_{0i}  \leq 13 A^{1/2} /(n_{i ({\rm cm}^3)}T_{i({\rm eV})})\label{Nph00}
\end{equation}

Photoelectron emission is thus expected to be of minor importance for water-group incident ions of density $n_i \geq 50$ cm$^{-3}$ and temperature $ \sim$ 1 eV. Likewise, "true" secondary electron emission is expected to be negligible, even for very small grains, for electron temperatures $ \sim$ 1 eV \citep{cho93}.

The above calculations  assume for simplicity that the particle velocity distributions are Maxwellian. Solar system plasmas are generally non-maxwellian, containing suprathermal particles with Kappa-like distributions  \citep{gar00,mey01,pie10}. This changes the fluxes  \citep{men94} and might have large consequences if the secondary electron emission were not negligible  \citep{mey82,cho93}. The grain relative velocity  has also been neglected, which is acceptable if it is smaller than the ion thermal speed.   

\subsection{Dusty environments \label{dusty}}

Finally, let us discuss briefly how the plasma electron depletion $n_e/n_i$ is related to the charged dust. Strictly speaking, the above calculations hold when the dust grains are \textquotedblleft isolated\textquotedblright, thus when  the grains' number density $n_d$ is small enough that their separation exceeds twice the Debye length, i.e.  $2 n_d^{1/3}L_D  < (3/4\pi)^{1/3}$. With the grains' charge  (\ref{Z1}) and the Debye length
\begin{equation}
L_D = [4\pi r_L (n_e+n_i)]^{-1/2} \label{LD}
\end{equation}
the ratio of the charge  carried by the grains to that available  in the medium is $|Z|n_d/(n_e+n_i)\simeq \eta P$ with
\begin{equation}
P \equiv 4 \pi  n_d a L_D^2  \label{sigma}
\end{equation}
Since  $P < (2 n_d^{1/3}L_D)^2 $  because the grains' separation necessarily exceeds their diameter, we have $P <1 $  if the grains' separation exceeeds $L_D$, so that in that case the grains' charge is not expected to perturb significantly the plasma.

\begin{figure}[t]
\noindent\includegraphics[width=110mm,clip]{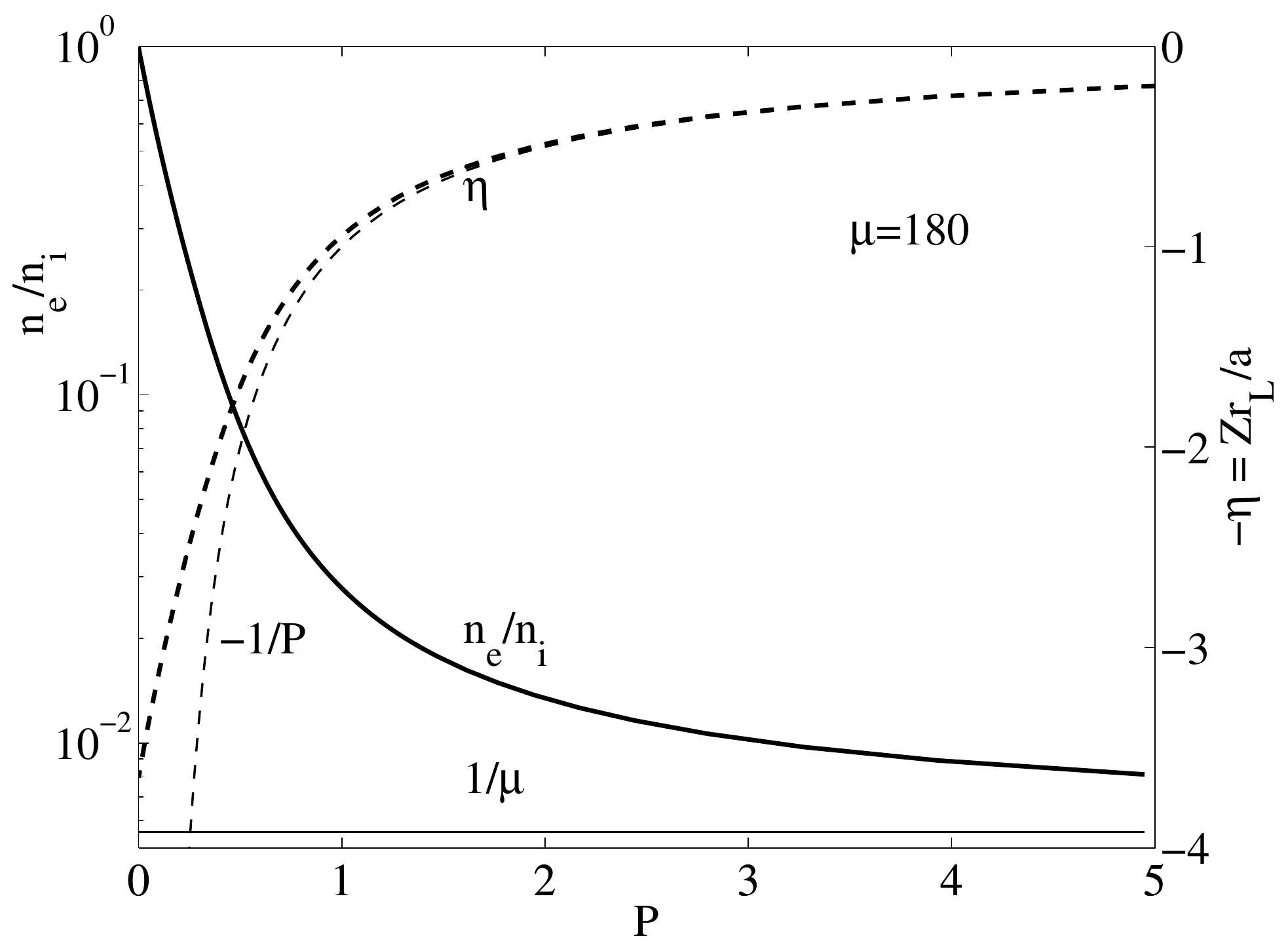}
\caption{ \label{fig1} Electron-to-ion number density $n_e/n_i$ (left axis, solid line) and normalized grain's charge $-\eta = Zr_L/a$ (right axis, dashed) versus the parameter $P = 4 \pi n_d \langle a\rangle L_D^2$,  from  (\ref{neni}) with $\mu = 180$ (water-group ions and electron sticking probability $s_e\simeq 1$) for dust grains of radius $a>r_L$. The  limits   $n_e/n_i \sim 1/\mu$  and $\eta \sim 1/P$  for $P\rightarrow \infty $  are plotted for comparison (thin lines).}
\end{figure}

In the opposite case   $P >1$, the grains' Debye spheres overlap and the electrons are depleted since many of them rest on the grain's surface, which in turn reduces the grain's charge   \citep{hav84,whi85}.  However, if the grain size is much smaller than the grains' separation, the increase of the grain-to-plasma capacitance due to the neighbouring grains can  be neglected and Eqs.(\ref{Nrep})-(\ref{Natt}) still hold \citep{whi85}, with $\eta \equiv -e\Phi /k_B T= -Zr_L/a$ where $\Phi$ is the grain-to-plasma potential.  Therefore   (\ref{Zeq}) holds in that case and using the  quasi-neutrality condition $n_i -n_e = - n_d Z $ and rewriting (\ref{LD})  as $n_i +n_e =  n_d \; a /P r_L$, one deduces straightforwardly
\begin{equation}
\frac{n_e}{n_i} = \frac{(1+\eta)e^{\eta}}{\mu}  =\frac{1-P \eta}{1+P \eta} \label{neni}
\end{equation}
From (\ref{neni}), one can deduce two of the three parameters $n_e/n_i$, $\eta $, and $P$, from one of them. Fig.~\ref{fig1} shows  $n_e/n_i$ versus $P$, which enables one to deduce the grains' properties from the electron depletion or vice-versa. For completeness we have also plotted $\eta$ versus $P$, first published by \cite{whi85} and reproduced in several papers.  Note that the pioneering results by  \cite{hav84,hav90} and references therein use a different definition of $P$, based on the plasma properties outside the dusty region - which is relevant for studying the local properties of a dust cloud; both definitions agree in the limit of small potentials (when the Boltzmann factors can be linearized).

In the limit $P \rightarrow \infty$, (\ref{neni}) yields
\begin{equation}
n_e/n_i  \simeq 1/ \mu   \;\;\;\;\;\;\;\; \eta  \simeq 1/ P \rightarrow 0 \label{ZZZ} 
\end{equation}
which corresponds to the limit $\mu n_e/n_i \rightarrow 1$ of  Eq.(\ref{Zeq}) and shows that the electrons cannot be more depleted with respect to ions than the limiting value $n_e/n_i  \simeq 1/ \mu$  \citep{men94}.

Of course, when the grains have  a continuous size distribution $dn_d/da$ (for  $a_{{\rm min}} <a<a_{{\rm max}}$),   $P$ must   be calculated by replacing in (\ref{sigma}) $a$ by its mean value $\langle a \rangle = \int _{a_{{\rm min}}}^{a_{{\rm max}}} da \; a(dn_d/da)/n_d$  \citep{hav90}. In general  the minimum $a_{{\rm min}}$ is unknown, but dust analyzers can measure the number density $n_d(a_0)$ of  grains larger than some radius $a_0>a_{{\rm min}}$. In that case, one can derive  $a_{{\rm min}}$ from the measured value of $n_e/n_i$ without having to make a hypothesis on the grains'  potential. For example, with  $dn_d/da\varpropto a^{-p}$  (with  $p>2$) so that the number density of grains larger than $a$ is $n_d(a) \varpropto a^{1-p}$, we substitute  $\langle a \rangle $ for $a$ in (\ref{sigma}) and  get
\begin{equation}
a_{{\rm min}}=\left(\frac{4 \pi L_D^2}{P}\frac{p-1}{p-2} \; n_d(a_0) \; a_0^{p-1}\right)^{1/(p-2)}   \label{amin}
\end{equation}
which together with (\ref{neni}) yields $a_{{\rm min}}$  as a function of $n_e/n_i$. This will be used in Sect.~\ref{enceladus}.

\section{Grains' polarization \label{image}}

The results of Sect.~\ref{chargingimpacts} hold for grains of size much larger than the Landau radius; how much larger will be determined later (Eq.(\ref{Zap}) and Fig.~\ref{fig3}).  For smaller grains, the polarization charges induced on a grain by approaching charged particles produce an electric potential which perturbs significantly their trajectories, as first shown in the context of ion capture by aerosols in the Earth's ionosphere \citep{nat60}; hence the particle fluxes are modified.

Consider an ion or an electron (charge $\pm e$) approaching at distance $r=ax$ from the centre of a spherical grain of radius $ a $ and electric charge $ Ze $. The approaching particle   is subjected to an electrostatic field that can be derived from the potential
\begin{equation}
\Phi (r)  =  \frac{e}{4\pi \epsilon _0 a}\left[\frac{Z}{x}\mp \frac{1}{2x^2(x^2-1)}\right] \;\;\;\;\;\;  x=r/a \label{Phir}
\end{equation}
obtained by adding to the Coulomb potential of the grain's charge that of the induced image  \citep{jac99}.

\subsection{Uncharged grains}

Consider first an uncharged grain ($Z=0$). Plasma particles  of charge $\pm e$ are subjected to the image term of the potential (\ref{Phir}), so that they are all attracted and their trajectories are bent towards the grain. Consider particles arriving isotropically from large distances at speed $v$. The impact parameter $p$ of the trajectory which barely grazes a sphere of radius $r$ is given by conservation of energy and momentum (in spherical coordinates) as
\begin{eqnarray}
p^2/a^2 &=& x^2+x_v/[2(x^2-1)]  \;\;\;\;\;\; x=r/a \label{p22} \\
{\rm where \;\;} \;\;\;\;\;\; x_{v} &=& [e^2/(4 \pi \epsilon_0 mv^2/2)]/a \label{xv}
\end{eqnarray}
For $r/a \rightarrow \infty$, (\ref{p22}) yields $p \rightarrow r$, as expected since the image potential becomes negligible at large distances and produces straight lines trajectories; at small distances, the trajectories are bent by the image force and for  $r/a \rightarrow 1$ we have $p/a \rightarrow \infty$ because of the increasing bending of the trajectories as the image term in (\ref{Phir})  increases. Between these two extremes, $p$ has a minimum $p_0$ obtained from  (\ref{p22}) by noting that $dp/dx = 0$ for $(x^2-1)^2 =x_{v}/2$, whence $
p^2_0/a^2 = 1+\sqrt{2x_{v}}  $. The particle random flux $snv/4$   is therefore increased by the factor $p^2_0/a^2$, which yields  the flux $N_v =  s n  \left[v+e\left(\pi \epsilon_0 m a\right)^{-1/2} \right]/4$. Averaging  over speeds and using $\langle v\rangle  = (8k_BT/\pi m)^{1/2}$  yields the flux
\begin{eqnarray}
N &=& N_0    F_0 \;\;\;\;\;\; {\rm for \;\;} Z=0 \label{Nuncharged}\\
F_0 &=&  1+ (\pi r_L/2a)^{1/2} \label{F0}
\end{eqnarray}
This exceeds the flux $N_0$ given by (\ref{N0}) by  the factor $F_0$, which may be very large when $a\ll r_L$.

\subsection{Repelling grains}

Now, consider particles which are repelled at distances $r \gg a$ (electrons if $Z<0$ or positive  ions if  $Z>0$). In that case, the two terms in the potential (\ref{Phir}) are of opposite signs so that the potential has a maximum at the distance $r_0$ given by
\begin{equation}
2x_0^2-1=|Z|\; x_0(x_0^2-1)^2 \;\;\;\;\;\;\;\;\;\;\;\;  x_0=r_0/a \label{x0}
\end{equation}
For $Z=1$ and 2, (\ref{x0}) yields  $x_0\simeq 1.62$ and 1.42 respectively, whereas in the limit $|Z| \rightarrow \infty $ we have  $x_0 \simeq  1+0.5 |Z|^{-1/2}$. At the maximum  of the barrier of potential, the potential energy of the charge $\pm e$ is
\begin{equation}
\pm e \Phi (r_0)  =  \frac{|Z|e^2}{4\pi \epsilon _0 a}\left[\frac{1}{x_0}- \frac{|Z|^{-1}}{2x_0^2(x_0^2-1)}\right] \label{Phir0}
\end{equation}
For  $Z \rightarrow +\infty $, the expansion of the bracket in  (\ref{Phir0}) to first order in $|Z|^{-1/2}$ yields $[1-|Z|^{-1/2}]$, whereas for $Z=1$ and 2 the bracket $\simeq 1/[1+|Z|^{-1/2}]$. Hence a reasonable approximation of (\ref{Phir0}) is
\begin{equation}
\pm e \Phi (r_0)  \simeq  \frac{1}{4\pi \epsilon _0 a}\; \frac{|Z|e^2}{1+|Z|^{-1/2}}   \label{Phir00}
\end{equation}

As soon as the approaching particles  come closer than $r_0$, they are attracted. Hence the effective collection radius is increased by a factor $y_0^2$, with $y_0 \simeq  x_0$ for $x_0 \gg 1$ since in that case the particles are weakly affected by the grain's charge farther than $r_0$. In the general case we have  $y_0<x_0$ because of the repelling electric force farther than $r_0$. Since only particles of kinetic energy exceeding  $|e \Phi (r_0)| $ can reach this distance, the flux of repelled particles with a Maxwellian distribution of temperature $T$ is given by
\begin{eqnarray}
N&=&N_0 y_0^2 e^{- |e \Phi(r_0)| /k_B T} \simeq N_0 F_r(Z)\;\;\;\;\;\;\; {\rm   repelled \; particles} \label{Nrep0} \\
F_r(Z)& \simeq & \left[ 1+(3 |Z|+4a/r_L)^{-1/2} \right]^2 e^{ -\left( \frac{|Z| r_L/a}{1+|Z| ^{-1/2}}\right)}  \label{Fr}
\end{eqnarray}
where we have used an approximation of $y_0 $ derived by \cite{dra87}.
Comparing with (\ref{Nrep}), one sees  that the polarization increases significantly the flux of repelled particles, by a factor     $\simeq 2.6 \times e^{ r_L/2a}$ for $|Z|=1$ and $a/r_L \ll 1$, which can be quite large for very small grains.

\subsection{Attracting grains}

Finally,  consider  particles which are attracted at distances $r \gg a$ (positive ions if $Z<0$ or electrons if  $Z>0$). In that case the focusing has two causes: first, the  field of the grain's charge $Ze$  which would yield the flux (\ref{Natt}) in the absence of polarization and acts far from the grain;  second, the  image contribution which  would yield the flux (\ref{Nuncharged}) when $Z=0$ and acts close to the grain. By comparing (\ref{Natt}) and (\ref{Nuncharged}), one sees that the attraction of the grain's charge  generally dominates, so that the flux is given by  (\ref{Natt}) with a correcting factor. Using an approximation for this factor   \citep{dra87}, one obtains  for a Maxwellian distribution at temperature $T$
\begin{eqnarray}
N &=& N_0 F_a(Z)  \;\;\;\;\;\; {\rm attracted \; particles} \label{Nattimage}\\
 F_a(Z)&=&\left(1+|Z| r_L/a\right) \left[ 1+ (|Z| +a/2r_L)^{-1/2} \right]  \label{Fa}
\end{eqnarray}

We conclude that when the grain's size does not exceed the Landau radius by a large amount, the polarization increases the fluxes whatever the grain's charge, thereby decreasing the charging time scales. Furthermore, because the flux of repelled particles is increased by a larger factor than the other ones because of the exponential term, the negative equilibrium charge tends to increase. However, since in that case $|Z|$ is not large, a statistical treatment of the grain charge distribution is needed.

\section{Charge probability distribution \label{proba}}

\subsection{Charge probabilities}

Let $f(Z)$ be the probability that a grain carries the charge $Ze$. The population of grains of charge  $Ze$ is depleted by collecting electrons and ions and replenished when grains of charge   $(Z+1)e$  collect electrons and when grains of charge   $(Z-1)e$  collect ions  (as discussed in Sect.~\ref{chargingimpacts}, we neglect secondary or  photoelectron emission).   Under stationary conditions,  this yields the simple recurrence relation \citep{dra87}
\begin{equation}
 f(Z)  N_i(Z) = f(Z+1)N_e(Z+1)   \label{balance}
\end{equation}
equivalent to a more complicated relation used by \citep{rap01}. Applying (\ref{balance}) iteratively with $N_e(Z)$ and  $N_i(Z)$  given by (\ref{Nuncharged}) for $Z=0$ and respectively by (\ref{Nrep0}) and (\ref{Nattimage}) for $Z<0 $ and the reverse for $Z>0$, we obtain
\begin{eqnarray}
f(Z)/f(0) &=& (\mu \; n_e/n_i)^{|Z|} F_0 \prod _{Z'=Z}^{-1}\left[\frac{F_r(Z'+1) }{F_a(Z')} \right]  \;\;\;\;\;\;\; Z<0 \label{fZ-} \\
f(Z)/f(0) &=& (\mu \; n_e/n_i)^{-Z}F_0 \prod _{Z'=1}^Z\left[\frac{F_r(Z'-1)}{F_a(Z')} \right] \;\;\;\;\;\;\; Z>0 \label{fZ+}
\end{eqnarray}
where $ F_0$, $ F_r$, $ F_a$ are defined respectively in (\ref{F0}), (\ref{Fr}), (\ref{Fa}), and we  set  $ F_r(0)=1$ in (\ref{fZ-}).  This can be solved by using $ \sum _{-\infty}^{+\infty} f(Z)=1$.

\subsection{Singly charged grains}

Eqs.(\ref{fZ-})-(\ref{fZ+}) yield in particular
\begin{eqnarray}
\frac{f(-1)}{f(0)}& \simeq & (\mu \; n_e/n_i) \frac{1+ (\pi r_L/2a)^{1/2}}{\left(1+ r_L/a\right) \left[ 1+ (1 +a/2r_L)^{-1/2} \right]} \label{f-1f0}\\
\frac{f(-1)}{f(+1)}& \simeq & (\mu \; n_e/n_i)^2 \label{f-1f+1}
\end{eqnarray}
Two important consequences emerge. First, since  $\mu \gg 1$  because of the large ion-to-electron mass ratio,  the positive grains are in minority  but they may be detectable, the more so as  the electrons are significantly depleted. Second, in the limit $a\ll r_L$, (\ref{f-1f0}) yields $f(-1)/f(0)\simeq (\mu \; n_e/n_i) \left(\pi  a /8r_L\right)^{1/2}$, an approximation which turns out to be  accurate to within $5 \%$ in the whole range $a/r_L \leq 1$.

\begin{figure}[t]
\noindent\includegraphics[width=110mm]{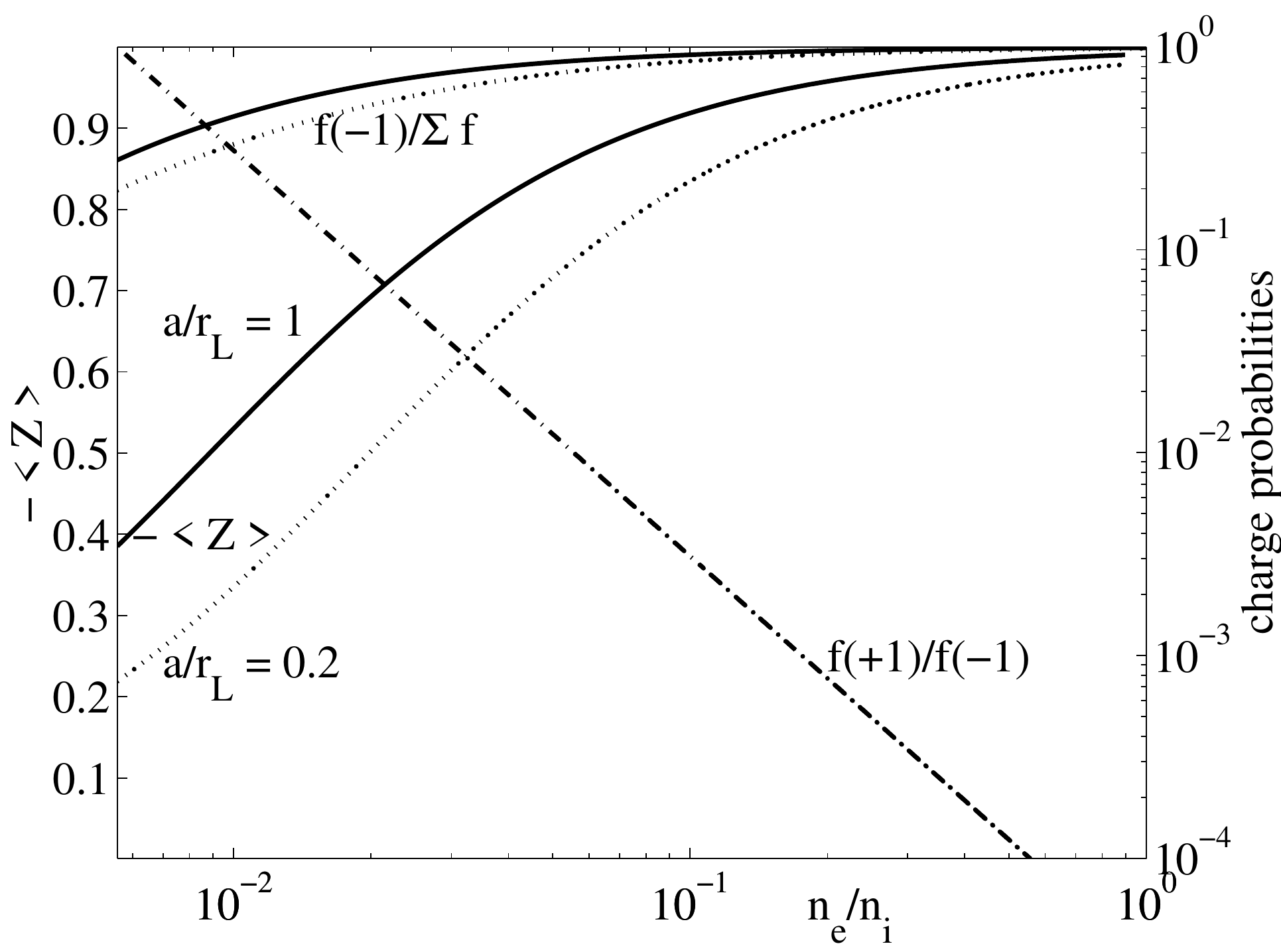}
\caption{ \label{fig2} Mean number of grain charge units $ \langle Z \rangle$  (left axis), proportion of negatively charged grains $f(-1)/\sum f$ (right axis) and ratio of  positively to negatively charged grains (dash-dotted, right axis) versus the electron-to-ion number density $n_e/n_i$ for two values of $a/ r_L$  with  $\mu = 180$ (water-group ions with $s_e \simeq 1$). These results concern grains of radius smaller than 1 nm for which the electron field emission limit (\ref{Zmax})  prevents multiple negative charging.}
\end{figure}

For small grains the probability is concentrated on the states  $Z=0$  and -1 since the number of negative charges is not only limited by the exponential factor in (\ref{Nrep0}), but also by electron field emission (e.g. \citep{men74,dra87}). Indeed,  for nanograins we have  $e^2/4\pi \epsilon _0 a \simeq 1.4$ eV, so that the surface Coulomb electric field  deforms significantly the potential barrier at the surface, which enables electrons inside to tunnel efficiently. In practice, this process becomes efficient when $|E|  >10^9$ V/m \citep{gom61}.  An ejected electron near the surface of a grain of charge $Ze<0$ will be subjected to the field amplitude $|E| \simeq (Z+1)e/4\pi \epsilon_0 a^2$. Hence the condition $|E|  <10^9$ V/m for electron field emission not to occur limits the grain charge state to
\begin{equation}
Z>-\left(1+0.7\: a_{nm}^2\right) \;\;\;\;\;\;\;{\rm Field \; emission \; limit} \label{Zmax}
\end{equation}
We do not consider ion field emission which limits the positive charging, since it requires a much higher field.

For $a\leq 1$ nm, electron field emission thus limits the (integer) number of grain charges $Q/e$ to $Z\geq-1$, as noted by \cite{hil12}. Hence,  the probability is concentrated on the states  $Z=0$  and -1, and the mean grains' charge number at equilibrium is $\langle Z\rangle \simeq -f(-1)= -1/[1+f(0)/f(-1)]$, i.e.
\begin{eqnarray}
\langle Z\rangle &\simeq & \frac{-1}{1+(8r_L/\pi a)^{1/2}/(\mu \; n_e/n_i)} \label{Zmean}\\
&\simeq & - \left[1+\frac{10^{-2} n_i/s_en_e}{(a_{{\rm nm}} \; T_{{\rm eV}})^{1/2}} \right]^{-1}  \label{Zmean1}
\end{eqnarray}
Therefore, the mean equilibrium charge of grains of radius $a \simeq 1$ nm  in a plasma of temperature $\simeq 1$ eV is roughly one electron if $s_e n_e/n_i \gg 10^{-2}$, and the ratio of positive to negative grains is according to (\ref{f-1f0})
\begin{equation}
f(+1)/f(-1) \simeq (5 \times 10^{-3} n_i/s_e n_e)^2  \label{B2}
\end{equation}

These values are plotted in Fig.~\ref{fig2}. They hold at equilibrium. The charging time scales  can be estimated from  the electron and ion flux on an uncharged grain, respectively, given from (\ref{Nuncharged})-(\ref{F0}), which yield
\begin{eqnarray}
\tau _{-1} &\simeq & \left[4\pi a^2 n_e s_e  \left(\frac{k_BT_e}{2\pi m_e}\right)^{1/2} \left(1+\left(\frac{\pi r_L}{2 a}\right)^{1/2}\right) \right]^{-1} \\
 &\simeq & \left[ 2 \times 10^{-6} s_e n_{e\;{\rm cm}^{-3}} \; a_{{\rm nm}}^2   \left( T_{{\rm eV}}^{1/2}+ \frac{1.5}{a_{{\rm nm}}  ^{1/2}} \right) \right]^{-1} \;\; {\rm s} \label{tau-}\\
\tau _{+1} &\simeq & (\mu n_e/n_i) \times \tau _{-1} \label{tau+}
\end{eqnarray}

\begin{figure}[t]
\noindent\includegraphics[width=110mm]{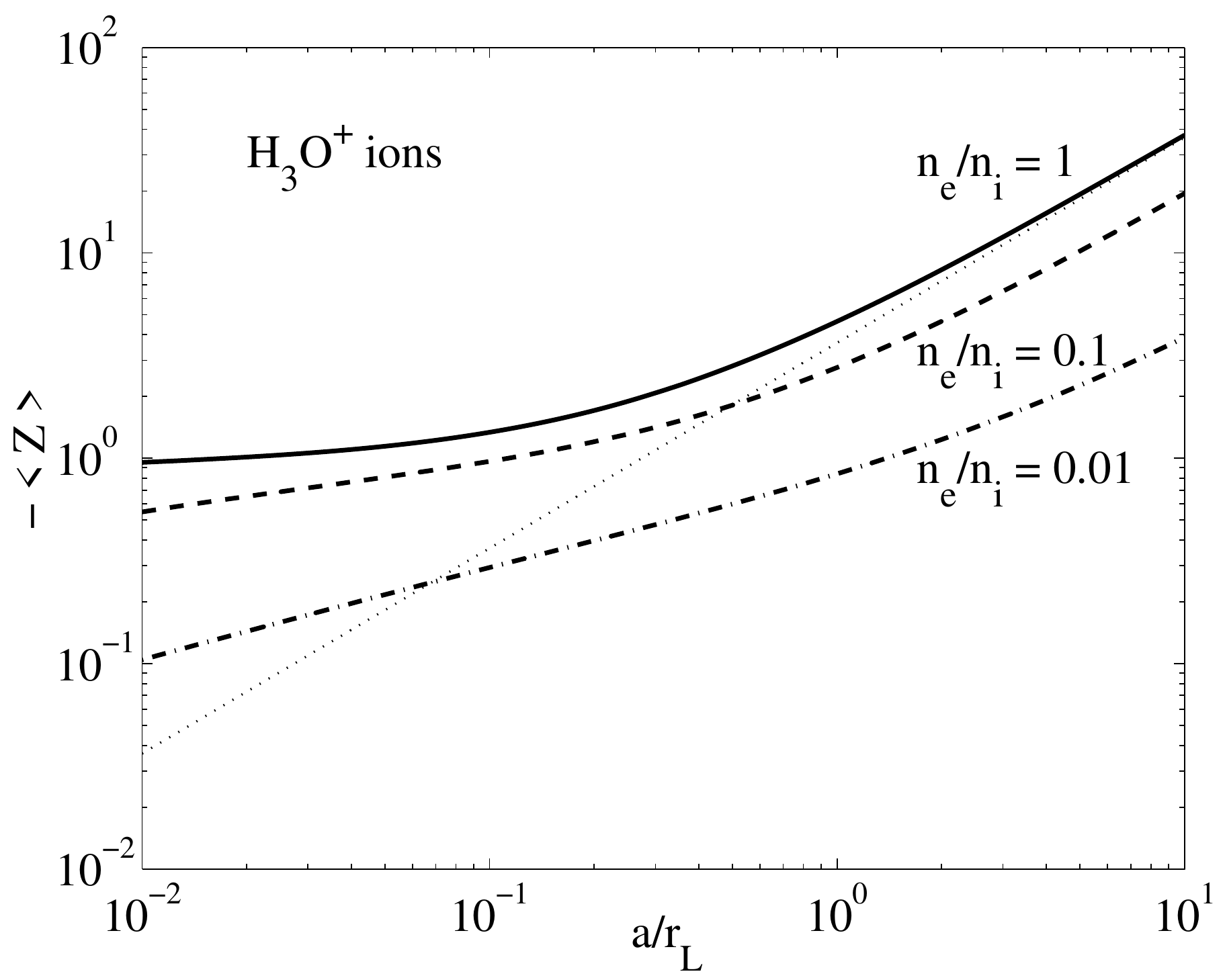}
\caption{ \label{fig3} Mean equilibrium number of electrons carried by a grain  from (\ref{Zap}) versus its normalized radius, for different values of the electron-to-ion number density   and  $\mu = 180$ (water-group ions with electron sticking coefficient $s_e\simeq 1$). The classical value  $-  \langle Z \rangle = 3.65 \; a/r_L$ is plotted (light dotted) for comparison. These values do not take into account the electron field emission limit, which  prevents multiple charging for grains of radius smaller than about 1 nm.}
\end{figure}

\subsection{Average charge of small or large grains}

Using (\ref{Zmean}) (valid for small grains) and (\ref{Z1}) with (\ref{Zeq}) (valid for large grains), we obtain the general approximation 
\begin{equation}
\langle Z\rangle \simeq  -\eta _0 \; a/r_L - \frac{1}{1+0.9\times 10^{-2}(n_i/n_e)(r_L/a)^{1/2}} \label{Zap}
\end{equation}
for water-group incident ions, with $\eta _0$ solution of  $e^{\eta _0}(1+\eta _0)\simeq 180\; n_e/n_i$  and $s_e=1$; if $s_e<1$, $n_e/n_i$ must be multiplied by $s_e$ in these expressions. A similar approximation was studied by \cite{dra87} in the special case $n_e=n_i$. Equation (\ref{Zap}) is plotted in Fig~\ref{fig3} for several values of $n_e/n_i$ and compared to the classical result valid for $a\gg r_L$ and $n_e \simeq n_i$ (dotted).

Two further consequences emerge. First,    the quasi-neutrality condition $n_e-n_i = n_d Z$ no longer yields (\ref{neni}) for grains of radius $a \lesssim r_L$ since  in that case  $\langle  Z\rangle $ is no longer given by the first term of (\ref{Zap}); one can nevertheless apply (\ref{neni}) in the Enceladus plume for larger grains, since they carry most of the dust total charge \citep{don12}. Second, the grains' charge-to-mass ratio, which governs their dynamics, varies faster with  mass for smaller grains. Indeed, whereas the charge-to-mass ratio of large grains varies in proportion of $Z/a^3\varpropto 1/a^2$ (from the dominant first term in (\ref{Zap})), the charge-to-mass ratio of smaller grains varies faster when the second term in (\ref{Zap}) is dominant;  if $n_e/n_i \simeq 10^{-2}$ this occurs as soon as $a\lesssim 10 \; r_L$, whereas if $n_e/n_i \simeq 1$, the charge-to-mass ratio  goes as $1/a^3$  for $a < r_L$.

\section{Enceladus nanograins \label{enceladus}}

Consider nanograins in the Enceladus plume, almost at rest  \citep{tok09} with respect to Enceladus as well as with respect to the plasma, of typical parameters:  $n_i \sim 10^3 - 3 \times 10^4$ cm$^{-3}$, $n_e/n_i \sim 10^{-2} - 10^{-1} $, $T \sim 1 $ eV  \citep{mor11,sha11},  main ion H$_3$O$^+$  \citep{cra09} and with a neutral gas density $n_0 \sim 5\times  10^7$ cm$^{-3}$ \citep{wai06}. This yields the Landau radius $r_L \simeq 1.4$ nm and the Debye length $L_D \simeq 0.07 \leftrightarrow 0.23$ m for $n_i \sim 10^4 \leftrightarrow 10^3 $ cm$^{-3}$ respectively.

\subsection{Negatively charged nanograins}

We deduce from (\ref{Zmean1}) (see also  Fig~\ref{fig2}) that more than 50\% of the grains  of radius  $a\simeq  1$ nm should carry one electron at equilibrium provided that $n_e /n_i > 10^{-2}/s_e$ ($s_e$ being the electron sticking probability), which confirms the value  inferred for these grains \citep{hil12}.

The corresponding charging time scale is given by (\ref{tau-}), which yields $\tau _{-1} \simeq 200 \leftrightarrow 2000 $ s for $n_e s_e\sim 10^3 \leftrightarrow 10^2 $  cm$^{-3}$ respectively. Note that neglecting the grains' polarization would yield charging time scales roughly twice larger than these values. With a speed $ \simeq 0.5$ km/s with respect to Enceladus - of the order of the gas bulk speed \citep{han08,don11}, this yields the free path for negative   charging $\simeq 0.4 \leftrightarrow 4 \; R_E$  (Enceladus radius $R_E \simeq 250$ km). This confirms the order-of-magnitude estimate by \cite{hil12} indicating that these nanograins are charged by the ambient plasma.

\subsection{Positively charged nanograins}

Consider now the  observed positively  charged grains \citep{jon09,hil12}, whose origin is under debate.  \cite{hil12} suggested three possible mechanisms: secondary electron emission, impacts of positive ions, and triboelectric charging (the latter first suggested by \cite{jon09}). The results of Sect.~\ref{proba} enable us to estimate the  grains' charging states resulting from the flux of ambient electrons and ions (taking into account the grains' polarization and charge discretization). According to  (\ref{B2}), we have  $f(+1)/f(-1) \sim (5 \times 10^{-3}n_i/s_en_e)^2 \sim 2 \times 10^{-1} \leftrightarrow 2 \times 10^{-3} $ for  $n_e/n_i \sim 10^{-2} \leftrightarrow  10^{-1} $  respectively (for $s_e \sim 1$). Since, according to (\ref{Zmax}),  grains of radius $a\simeq 1$ nm or smaller  do not carry more than one electron, $f(+1)/f(-1$) yields the ratio of positively to negatively charged  grains. However this is an equilibrium value, which holds when the time involved exceeds the larger charging time scale $\tau _{+1}$; otherwise, the ratio $f(+1)/f(-1) $ should be smaller since  $\tau _{+1}>\tau _{-1}$. With   $n_i \simeq 3 \times 10^4$ cm$^{-3}$,  (\ref{tau+}) yields $ \tau _{+1} \simeq 1300 $ s. With a speed $ \simeq 0.5$ km/s, this yields a  free path for (positive) charging $\simeq 2  R_E$. This value is of the order of the involved paths, which suggests that a significant proportion of these nanograins have their equilibrium charge. These estimates are a strong indication that the impacts of ambient plasma particles can explain both the negatively and positively charged grains, without having to rely on other charging processes.

\subsection{Electron depletion}

The charge of these nanograins   cannot compensate for the strong observed electron depletion, and  larger grains act to achieve plasma quasi-neutrality \citep{yar09,far10}. When those  \textquotedblleft large grains\textquotedblright   represent the major contribution to  the total dust charge,   Fig.~\ref{fig1} (Eq.(\ref{neni})) enables one to deduce directly their properties from the observed electron depletion or vice-versa, via $P$. For example, let us assume  $dn_d/da \varpropto a^{-p}$ for $a>a_{min}$ with $p=4.5$ and a number density $n_d(a_0)\simeq 0.1$ cm$^{-3}$ for grains of radius $a_0>2 \; \mu $m, according to typical measurements \citep{spa06,kem08}.   Eqs.(\ref{neni}) and (\ref{amin}) show that with an observed electron depletion $n_e/n_i\simeq 10^{-2}$, quasi-neutrality can be achieved with $a_{min} \simeq 0.1 \; \mu $m. Note that this estimate does not require the grain  potential as an input since it is calculated in parallel,  contrary to estimates using the measured spacecraft potential (e.g. \cite{sha11}).

Finally, according to Eq.(\ref{ZZZ}), the electrons cannot be more depleted than $ n_e/n_i \simeq 1/\mu \simeq 5 \times 10^{-3}/s_e \geq  5 \times 10^{-3}$ (since $s_e\leq 1$),  which agrees in order of magnitude with observation.

\subsection{Minimum size of nanograins}

Finally, consider grains of radius  smaller than about 1  nm. According to (\ref{Zmean1}), their charge should decrease  below one charge unit when $a_{{\rm nm}}<(10^{-2}n_i/s_en_e)^2/T_{{\rm eV}}$,  whereas  their charging time scale increases. This might possibly contribute to  the decrease observed  in the flux of charged  grains below 1 nm \citep{hil12,jon12}.

Several other effects are expected to act at such small sizes. First, the high electric field $E_0 = Ze/4\pi \epsilon _0 a^2$ at the grain's surface can make it explode if the electrostatic stress $\epsilon _0 E_0^2$ exceeds the maximum grain's tensile strength against fracture $S$ (e.g. \citep{hil79,dra79}). The condition $\epsilon _0 E_0^2 <S$  yields the limiting grain radius
\begin{equation}
a_{{\rm nm}} > 0.36 \times |Z|^{1/2} (S/10^9 \; {\rm Nm}^{-2} )^{-1/4} \label{fracture}
\end{equation}
The maximum tensile strength $S$ of sub-nanometric ice grains is highly uncertain. Since the tensile strength of macroscopic materials is  determined in a large part by cracks and dislocations, $S$ is expected to exceed by a large amount the (highly temperature dependent) value of macroscopic ice $S \sim 10^7$ Nm$^{-2}$ \citep{cro79}, for sub-nanometric grains having a compact structure.

Given these uncertainties and the weak dependence on $S$ of the size limit (\ref{fracture}), we make below a tentative order-of-magnitude estimate. With the   bonding strength  energy $\simeq 0.8 \times 10^{-2}$ eV per hydrogen bond, 4 hydrogen bonds per water molecule, and assuming bulk water ice density i.e. about $ 3 \times 10^{28}$ water molecules/m$^3$, we obtain $\sim 2 \times 10^8$ Jm$^{-3}$ (equivalent to force per unit area), from which we deduce the  tentative strength against fracture $S \sim 2 \times 10^8$ Nm$^{-2}$.

Substituting this value in (\ref{fracture}) with $Z=1$ yields  the minimum grain radius $a_{{\rm min}} \simeq  0.7$ nm. This figure varies with the grain's tensile strength $S$ as $S^{-1/4}$, so that varying $S$ by a factor of 5 would produce a variation in $a_{{\rm min}}$ of about 50 \%. Note also that 0.7 nm is roughly the size of a unit cell of a Ih ice crystal and twice the width of an elementary step of ice crystallization \citep{saz10}.

Another size limitation might be produced by the centrifugal stress due to  the grain's spin induced by impacts of molecules \citep{spi78,dra79}. Contrary to the above electrostatic limit, it also acts  on  uncharged grains. At equilibrium (justified by Eq.~(\ref{tauspin})), the rms angular speed $\omega $ of a grain due to collisions with neutrals of temperature $T_0$ satisfies $I\omega ^2 \simeq 3 k_B T_0$ where $I \simeq (8/15)\pi \rho a^5$. A spinning grain of mass density $\rho$ will be destroyed if $(\pi/8)\rho a^2 \omega ^2 > S$, which yields the survival condition  (e.g. \citep{mey84})
\begin{equation}
a_{{\rm nm}} > 0.5 \times T_{0\;{\rm eV}}^{1/3}  (S/10^9 \; {\rm Nm}^{-2} )^{-1/3} \label{spin}
\end{equation}
Substituting   $T_0 \simeq  0.02 $ eV \citep{wai06} and the order of magnitude  $S \sim 2 \times 10^8$ Nm$^{-2}$ determined above, we obtain  $a > 0.2$ nm. Since this limit is smaller than the electrostatic disruption limit  (\ref{fracture}), centrifugal disruption is not expected to play a major role.  Note that we do not consider the spin induced by the impacts of ions of temperature $T$ (although this would yield a size limit higher by the factor $(T/T_0)^{1/3}$), because the ion number density is too small for inducing  a significant grain's spin during the time scales involved. Indeed, for the grains to acquire a spin governed by the thermal energy of a particle species,  they should  have been struck by their own mass of these particles. Applying (\ref{N0}) to H$_2$O molecules of mass $m_0 \simeq 18 \; m_p$  yields the time scale
\begin{equation}
\tau_{{\rm spin}} \simeq \frac{(2\pi)^{1/2}\rho a}{3n_0 (m_0 k_B T_0)^{1/2}} \sim  10^{10} a_{{\rm nm}}/(n_{0 \;{\rm cm}^{-3} } T_{0\;{\rm eV}}^{1/2}) \;\;{\rm s} \label{tauspin}
\end{equation}
whence  $\tau_{{\rm spin}} \sim 1.5 \times 10^3$ s,  which is of the order of magnitude of the time scales involved and  smaller by more than two orders of magnitude than the value for ion impacts (which confirms that the latter do not affect the grains' spin).

\section{Concluding remarks}

We have derived analytical expressions for the charge of nanograins in cold dense and dusty environments, under conditions relevant in the outer solar system, and applied them to Enceladus nanograins. This analysis shows  that a large proportion of nanograins should be charged with one electron, as assumed in previous studies and argued by \cite{hil12}, and  that the impacts of ambient ions should explain the observed positively charged grains without having to assume other charging processes. Electrostatic stresses are expected to limit the size of charged grains to a minimum radius of about 0.7  nm - a value which should be taken with caution since it assumes a compact structure and varies with the badly known grain tensile strength $S$ in proportion of  $S^{-1/4}$. This effect might contribute to  the  strong  decrease in the grain number density observed at radii below about 1 nm \citep{jon09,hil12}.

However, subnanometric ice grains fall into the uncertain transition region between macroscopic and microscopic behavior. In particular, the electron sticking coefficient $s_e$ is expected to decrease nearly linearly with radius for $a \lesssim 1$ nm \citep{vos06,meg09}; according to (\ref{f-1f0}) and (\ref{Zmean1}), this should decrease the grain's charge and therefore the probability of detection below $\simeq 1$ nm.

Furthermore, the detailed physics of ice crystallization is still not understood, and a grain of radius 1 nm with the density of ice contains only $\simeq 140 $ H$_2$O molecules, more than 50 \% of which should lie at the surface. This number  is smaller than the recently determined minimum number of $ 250-300$ H$_2$O molecules required for  crystallization of a water cluster  \citep{pra12}, which corresponds to a radius of about 1.3 $\mu$m with the density of ice. This radius of 1.3 $\mu$m (whose dependence on temperature is unknown and which should be taken with caution in the context of Enceladus plume),  is close to the observed onset of flux decrease  \citep{jon09,hil12}.

\section*{Acknowledgments}
I thank Tom Hill and another reviewer for their helpful comments on the manuscript.

\newpage

\end{document}